\documentclass[11pt,a4paper]{article}
\usepackage[T1]{fontenc}
\usepackage[utf8]{inputenc}
\usepackage{lmodern}
\usepackage[margin=2.5cm]{geometry}
\usepackage{amsmath,amssymb,bm,graphicx}
\usepackage[labelfont=bf,font=small,labelsep=quad]{caption}
\usepackage[hidelinks]{hyperref}

\hypersetup{pdftitle={LieGROG: A Lie algebra approach for calibrating GRAPPA operators},
  pdfauthor={Tobias Wech},
  pdfsubject={Preprint, submitted to Magnetic Resonance in Medicine},
  pdfkeywords={non-Cartesian, GRAPPA operator gridding, GROG, self-calibration, Toeplitz-embedding}}
\graphicspath{{Figures/}}

\newcommand{\Gop}{\mathbf{G}}
\newcommand{\sv}{\mathbf{s}}
\newcommand{\kv}{\mathbf{k}}
\newcommand{\Lop}{\mathbf{L}}
\newcommand{\Em}{\mathbf{E}}
\newcommand{\NC}{N_{\mathrm{C}}}
\newcommand{\dk}{\Delta\kv}
\newcommand{\Am}{\mathbf{A}}
\newcommand{\Wm}{\mathbf{W}}
\newcommand{\GCart}{\mathrm{CartGROG}}

\begin{document}

\begin{center}
{\small Preprint. Submitted to \emph{Magnetic Resonance in Medicine}.}\par\vspace{1.5em}
{\LARGE\bfseries LieGROG: A Lie algebra approach for calibrating GRAPPA operators\par}\vspace{1.2em}
{\large Tobias Wech$^{1,2}$\par}\vspace{0.8em}
{\small $^1$Department of Diagnostic and Interventional Radiology, University Hospital W\"urzburg, Germany\\
$^2$Comprehensive Heart Failure Center, University Hospital W\"urzburg, Germany\par}
\end{center}

\noindent{\small\textbf{Correspondence:} Tobias Wech, Department of Diagnostic and Interventional Radiology, University Hospital W\"urzburg, Oberd\"urrbacher Str. 6, 97080 W\"urzburg, Germany. Email: \href{mailto:wech_t@ukw.de}{wech\_t@ukw.de}\par}

\begin{abstract}
\setlength{\parindent}{0pt}\setlength{\parskip}{0.4em}
\textbf{Purpose:} In multi-coil acquisitions based on non-Cartesian k-space sampling, image reconstruction using GRAPPA Operator based gridding (GROG) can provide advantages with respect to the non-uniform FFT (NUFFT): 
It waives both the need for calculating a density compensation function, and for transferring back and forth between Cartesian and non-Cartesian points in iterative algorithms.
Non-trivial noise propagation and self-calibration limited to specific trajectories, however, have contributed to GROG being rather sparsely applied so far.

\textbf{Theory and Methods:} The group structure of the GRAPPA Operator allows estimating its generator as a single joint fit for all k-space dimensions within the associated Lie algebra. Assuming a linear regime, 
the generator can be fitted using subsequent pairs of sampling points from nearly any arbitrary non-Cartesian read-out.
This approach called LieGROG was evaluated against classical GROG calibration and NUFFT/Toeplitz, using simulations with an analytic ground truth and various 2D and 3D MR-scans employing typical non-Cartesian trajectories.

\textbf{Results:} Throughout all applications, LieGROG proved to provide fast, robust and straightforward calibration of the GRAPPA Operator. It clearly outperformed the classical GROG calibration approaches in terms of noise amplification. 
LieGROG matched the results of the NUFFT/Toeplitz-based reference MR reconstructions closely, while being 6 to 189 times faster and reducing peak VRAM by 20 to 45\,\% in the 3D experiments.

\textbf{Conclusion:} LieGROG transforms GROG into a more versatile and stable tool, ultimately aligning non-Cartesian imaging with the standard, FFT-based reconstruction pipelines established for Cartesian scans.
\end{abstract}


\section{Introduction}

Compared to standard Cartesian scanning, non-Cartesian k-space trajectories have several advantages, such as increased robustness against motion, improved scan-efficiency and a general higher flexibility. 
However, it also renders image reconstruction more complex, as data have to be transferred to a regular grid prior to the application of the fast Fourier Transform (FFT).
The standard solution is convolution gridding and the associated non-uniform FFT (NUFFT \cite{Fessler2003,Jackson1991}), which in brief grids onto the Cartesian lattice by convolving with a (Kaiser-Bessel) kernel, corrects sampling density by an estimated function (density compensation function, DCF), 
applies an oversampled inverse FFT and finally an apodization correction.

Various efficient open-source implementations of the NUFFT have been presented \cite{Fessler2004,BART,Ong2019,Muckley2020} which partly also provide GPU acceleration. For reconstructions based on a single 2D transform from k- to image-space, computing capacity of a standard PC is typically sufficient. 
For 3D acquisitions (grid size $N_x \times N_y \times N_z$) with $\NC$ coils and an oversampling factor of $\alpha$, however, VRAM capabilities of common GPUs can very well be exceeded for typical grid parameters, as the peak memory scales as $O(\NC\,\alpha^3 N_x N_y N_z)$. In iterative reconstruction 
approaches like CG-SENSE \cite{Pruessmann2001, Maier2021}, Compressed Sensing \cite{Lustig2007} or Variational Networks \cite{Hammernik2018, Kleineisel2022} the multi-coil data $\mathbf{m}$ are modeled by applying an encoding operator $\Am$ to the image $\mathbf{v}$, 
which weights the image with the coil sensitivities $C_j$ and transforms it to the sampling positions of the k-space trajectory. CG-SENSE solves the normal equations \cite{Pruessmann2001}
\begin{equation}
  \Am^{H}\Wm\Am\,\mathbf{v} \;=\; \Am^{H}\Wm\,\mathbf{m},
  \label{eq:normaleq}
\end{equation}
where the diagonal $\Wm$ weights the acquired samples, e.g. by a DCF. The necessity of transferring data to the Cartesian grid and back in each iteration can easily represent the main bottleneck in terms of reconstruction speed, especially in 3D.

To mitigate this, Toeplitz embedding \cite{Fessler2004, Wajer2001} can be exploited: The repetitive application of $\Am^{H}\Wm\Am$ within the iterations is replaced by a computationally cheaper convolution with a kernel, which can be pre-computed based on the k-space trajectory and $\Wm$,
the coil sensitivities remaining outside it. The right-hand side of Eq.~\eqref{eq:normaleq}, however, still requires one application of $\Am^{H}$ to the acquired data, which no kernel can replace. For noise-optimal reconstructions the $\Wm$ is the identity \cite{Pruessmann2001,Maier2021}. 
Using the DCF in $\Wm$ instead, allows accelerating the convergence, at the price of altering the estimator \cite{Pruessmann2001}.

Alternatively, when acquiring MR data with a coil-array, the encoding capacity of the array itself can be exploited to shift off-grid measurements to the Cartesian grid. This was proposed as GRAPPA Operator Gridding (GROG \cite{Griswold2005,Seiberlich2007,Seiberlich2008}),
which allows a single initial transformation of a non-Cartesian into a Cartesian reconstruction problem, with all the accompanying advantages for subsequent (iterative) modules. 
Furthermore, it intrinsically corrects for the sampling density, such that no DCF has to be estimated. 
Noise amplification and true self-calibration only for specific trajectories (predominantly 2D radial) have, however, most likely contributed to GROG still rather being a niche tool in research \cite{Benkert2018, Schad2026}.

In this paper, a new method called "LieGROG" is proposed, which aims at providing a more flexible, fast and robust calibration of the GRAPPA operator, thereby bringing the processability of non-Cartesian scans close to Cartesian ones.

\section{Theory}\label{sec:recap}

This chapter first recaps the GRAPPA operator and GROG, followed by a discussion of current limitations and finally introduces LieGROG.

\subsection{The GRAPPA operator.}\label{sec:propagator}

After introducing GRAPPA \cite{Griswold2002} in 2002, Griswold and colleagues extended the theory in 2005 \cite{Griswold2005} by proposing an elegant operator formalism:
Given the multi-coil signal of a complex-valued coil vector $\sv(\kv)$ of length $\NC$ at position $\kv$, the signal at a shifted position $\kv+\dk$ can be calculated by applying an $[\NC\!\times\!\NC]$ matrix $\Gop(\dk)$ holding the GRAPPA weights for a single source and a single target point separated by the displacement vector $\dk$:
\begin{equation}
  \sv(\kv+\dk)=\Gop(\dk)\,\sv(\kv),\quad \Gop(\dk)\in\mathbb{C}^{\NC\times\NC}.
  \label{eq:gop}
\end{equation}
The authors pointed out the similarity between $\Gop$ and the propagator operator in quantum mechanics \cite{Griswold2005}, i.e. the time-evolution operator $U(t)$ \cite{Sakurai2020}, which for a time-independent Hamiltonian $\hat H$
reads:
\begin{equation}
  U(t)=\exp\!\Big(-\tfrac{i}{\hbar}\hat H\,t\Big),\qquad U(s)\,U(t)=U(s{+}t).
  \label{eq:propagator}
\end{equation}
$U(t)$ forms a one-parameter \emph{Lie group} \cite{Hall} and $\hat H$ is its \emph{generator}, representing the total energy of the system and driving the smooth transition between quantum mechanical states. $\hbar$ is the reduced Planck constant.

Strictly, the analogy only holds in an idealized setting, where \emph{any arbitrary} k-space shift $\delta$ can be delivered the fractional power $\Gop^{\,\delta}$. In this case, $\Gop$ also forms a one parameter Lie group:
Knowing a single coil vector $\sv(\kv)$ (``initial state'') would define the entire k-space, and thus the image. 
It intrinsically assumes, however, that the underlying coil sensitivities span a complete harmonic Fourier basis over the field of view (FOV), with as many independent modes as image pixels to be resolved in a certain dimension.
In typical clinical scenarios, $\NC$ is in the order of 10-40 and the aimed depiction of the imaged object has clearly more degrees of freedom. 
However, the operator still "knows" the multi-coil signal within a small vicinity here, whose size depends on the object and the coil array.

\subsection{GRAPPA Operator Gridding --- GROG}

In the original publications on GROG \cite{Seiberlich2007}, Seiberlich and colleagues exploited this feature for gridding non-Cartesian MR scans: GRAPPA operators were first calibrated separately for each dimension using a Cartesian calibration scan. Subsequently, they were used to shift the data, point by point, 
from the non-Cartesian positions to the nearest Cartesian ones:
\begin{equation}
  \begin{aligned}
    &\sv(k_x+\delta_x,k_y+\delta_y,k_z+\delta_z)\\
    &\qquad=\;\Gop_x^{\,\delta_x}\,\Gop_y^{\,\delta_y}\,\Gop_z^{\,\delta_z}\,\sv(k_x,k_y,k_z)
  \end{aligned}
  \label{eq:gfactor}
\end{equation}
If a Cartesian grid point was the target of more than just one sample, the entry was built by averaging. This intrinsically compensates for unequal sampling densities across k-space with no need for estimating any DCF.
\begin{figure}[!htb]\centering
  \includegraphics[width=\linewidth]{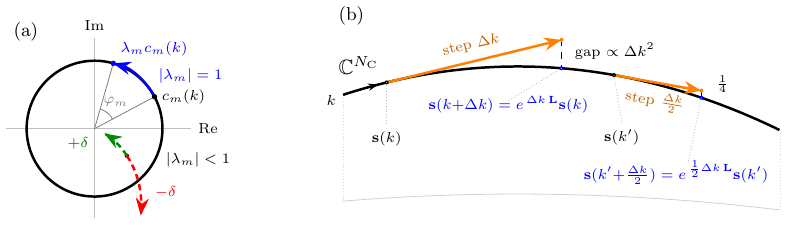}
  \caption{\textbf{Using GRAPPA operators for ``traveling the k-space orbit'' of an image.}
  \textbf{(a)} Projection of the orbit, i.e. the multi-coil signal $\sv(\kv)$ parametrized by the k-space position $\kv$, onto the eigenmodes: 
  An ideal GRAPPA operator features $|\lambda_m|=1$ for all $m$ and only ``rotates'' multi-coil data in this view. Limited encoding capacity of 
  real-world arrays, noise in the calibration data and the regularization of the fit can result in eigenvalues no longer being on the unit circle. 
  A $|\lambda_m|$ close to zero thereby leads to a strong amplification of the noise encoded in the corresponding eigenvector in backward shifts ($-\delta$).
  \textbf{(b)} The orbit in coil space $\mathbb{C}^{\NC}$: $\Gop$ acts like a time-evolution operator which evolves $\sv(\kv)$ along the k-space trajectory. 
  The tangent vector to the orbit at $\sv(\kv)$ is $\mathrm{d}\sv/\mathrm{d}k=\Lop\,\sv(\kv)$, with the generator $\Lop$ an element of the Lie algebra.
  The closer consecutive readout points lie on the orbit (i.e. the smaller $\Delta k$), the better this tangent approximates the true path. 
  In this regime of small $\Delta k$, all components of $\Lop$ are jointly fitted from the measured data, thereby calibrating $\Gop$.}
  \label{fig:genexp}
\end{figure}

The fractional power of the GRAPPA operator, i.e. the ``small shift operator'' $\Gop^{\,\delta}$ of \cite{Seiberlich2007} for sub-cell shifts $|\delta|<1$, can be calculated in its eigenbasis \cite{Seiberlich2007}:
 \begin{equation}
  \begin{gathered}
    \Gop^{\,\delta}\;=\;\Em\,\bm\Lambda^{\delta}\,\Em^{-1},\\
    \bm\Lambda^{\delta}\;=\;\operatorname{diag}\!\big(\lambda_1^{\delta},\dots,\lambda_{\NC}^{\delta}\big),\\
    \lambda^{\delta}\;:=\;e^{\delta\ln\lambda},
  \end{gathered}
   \label{eq:frac}
 \end{equation}
by simply modulating the eigenvalues $\lambda_m \rightarrow \lambda_m^{\delta}$. $\ln$ denotes the principal branch, $\arg\lambda\in(-\pi,\pi]$.

An ideal shift in k-space now corresponds to multiplying the multi-coil signal with a linear phase ramp in image space, which touches the phase only, not the magnitude, i.e.$|\lambda_m|=1$ for all $m$ (Fig.~\ref{fig:genexp}a).
A real array with $\NC$ channels can not generally represent this ramp exactly, and the best $\Gop$ which can be expressed by the array both rotates and attenuates ($|\lambda_m|\le1$).
The corresponding deviation from the ideal unit circle (Fig.~\ref{fig:genexp}a) is the minimum error achievable for a given coil setup.
For realistic arrays, the shrinkage is moderate, and shifts below half a cell move $|\lambda_m^{\delta}|$ back towards unity. Any error beyond this inevitable ``array floor'' originates from the calibration of $\Gop$.

\paragraph{Cartesian calibration.}
In \cite{Seiberlich2007}, a separate GRAPPA operator $\Gop_a$ is estimated for each axis $a$ ($a = x,y,z$). For a Cartesian reference scan with consecutive samplings spaced by $\mathbf e_a$, many readout pairs $\{\sv(\kv),\sv(\kv+\mathbf e_a)\}$ can be built and arranged as follows:
\begin{equation}
  \begin{gathered}
    S_0=[\,\sv(\kv_1)\;\cdots\;\sv(\kv_M)\,],\\
    S_1=[\,\sv(\kv_1{+}\mathbf e_a)\;\cdots\;\sv(\kv_M{+}\mathbf e_a)\,]
  \end{gathered}
  \label{eq:fitG}
\end{equation}
Eq.~\eqref{eq:gop} delivers $S_1=\Gop_a S_0$. For $M \gg \NC$ this system is overdetermined. Due to the \emph{array floor} and measurement noise, no exact solution exists. Thus, $\Gop_a$ is determined in a least-squares manner, i.e. by determining the pseudoinverse \cite{Seiberlich2007} 
\begin{equation}
  \Gop_a \;=\; S_1\, \mathrm{pinv}(S_0).
\end{equation}
For real arrays, the set $C_j$ can be \emph{nearly} linearly dependent, i.e. linear combinations of $C_j$ nearly vanish across the object, 
which renders the Gram matrix $S_0S_0^{H}$ ill-conditioned. In this case, its singular vectors with the smallest singular values are dominated by noise, which 
can be managed by Tikhonov regularization in practice \cite{pygrappa}:
\begin{equation}
\begin{aligned}
  &\Gop_a \;=\; \big(S_1 S_0^{\!H}\big)\big(S_0 S_0^{\!H}+\beta\,I\big)^{-1},\\[4pt]
  &\qquad \beta\;=\;\beta_{\mathrm{rel}}\,\frac{\operatorname{tr}\big(S_0S_0^{H}\big)}{\NC}.
\end{aligned}
  \label{eq:fitGreg}
\end{equation}
The penalty $\beta$ damps the contributions of small singular values, which would otherwise transfer measurement noise into the fit. 
Here, $\beta$ is split into a dimensionless $\beta_{\mathrm{rel}}$ and the mean diagonal entry of the Gram matrix, such that one choice of $\beta_{\mathrm{rel}}$ behaves the same across matrix sizes, number of coils and signals.  

For a weakly conditioned Gram matrix, both the noise within $S_0$ and the regularization by $\beta$ draw the eigenvalues of ``weak directions'' in $\Gop_a$ towards zero, adding to the shrinkage caused by the array floor.
While positive shifts according to Eq.~\eqref{eq:frac} are rather benign, negative ones ($\delta < 0$, typically $\sim$50\,\% of the shifts) lead to noise amplification (Fig.~\ref{fig:genexp}a): $|\lambda_m^{\delta}| \gg 1$ for $|\lambda_m| \approx 0$ and the coefficients of corresponding eigenvectors $\Em_m$ are dominated by noise.

If no Cartesian reference is available, but Nyquist sampled k-space regions can be built out of the non-Cartesian data, NUFFT-based gridding can be used to synthesize Cartesian calibration data \cite{Hamilton2023,Schad2026}.   

This family of GROG calibrations will be referred to as $\GCart$.

\paragraph{Self-calibrating GROG (SC-GROG).}\label{sec:scgrog}
A self-calibrated version of GROG (SC-GROG), which is mainly practicable for radial trajectories, was proposed in \cite{Seiberlich2008}: In a first step, individual $\Gop_a = \Gop_\theta$ 
are fitted in accordance with $\GCart$, now separately for each radial direction $\theta$ acquired (outlined for regular 2D radial here, see \cite{Seiberlich2008} for details):
\begin{equation}
  \begin{gathered}
   \Gop_\theta \;=\; S_1\, \mathrm{pinv}(S_0), \\ %
    S_0=[\,\sv(\kv_1)\;\cdots\;\sv(\kv_M)\,],\\
    S_1=[\,\sv(\kv_1{+}\mathbf e_\theta)\;\cdots\;\sv(\kv_M{+}\mathbf e_\theta)\,]
  \end{gathered}
  \label{eq:Gtheta}
\end{equation}
Assuming $\Gop_\theta=\Gop_x^{\,\delta_x}\Gop_y^{\,\delta_y}$ \cite{Griswold2005,Seiberlich2008} and taking the matrix logarithm, delivers the linearized equation
\begin{equation}
  \ln\Gop_{\theta} \;=\; \delta_x\,\ln\Gop_x \;+\; \delta_y\,\ln\Gop_y.
  \label{eq:logsystem}
\end{equation}
For $N_{\mathrm{proj}}$ directions $\theta_i$ the steps $(\delta_x,\delta_y)_i$ form a distance matrix $[\bm\delta_x,\bm\delta_y]$ of size
$N_{\mathrm{proj}}\times2$. Each entry $(p,q)$ of the $\NC^2$-sized matrices $\Gop_x$ and $\Gop_y$ then follows
from its own separate least-squares problem
\begin{equation}
\begin{aligned}
  &\big[\ln\Gop_{\theta_1}(p,q),\dots,\ln\Gop_{\theta_{N_{\mathrm{proj}}}}(p,q)\big]^{T} \;=\; \\
  &\qquad \qquad [\bm\delta_x,\bm\delta_y]\,\big[\ln\Gop_x(p,q),\,\ln\Gop_y(p,q)\big]^{T},\\[10pt]
  &\big[\ln\Gop_x(p,q),\,\ln\Gop_y(p,q)\big]^{T} \;=\; \\
  & \enspace \mathrm{pinv}\big([\bm\delta_x,\bm\delta_y]\big)\,
     \big[\ln\Gop_{\theta_1}(p,q),\dots,\ln\Gop_{\theta_{N_{\mathrm{proj}}}}(p,q)\big]^{T}
\end{aligned}
  \label{eq:sep}
\end{equation}
and a final matrix exponential
\begin{equation}
\begin{aligned}
  \Gop_x=\exp(\ln\Gop_x),\quad &\Gop_y=\exp(\ln\Gop_y).
\end{aligned}
  \label{eq:sep2}
\end{equation}

The exponential form Eq.~\eqref{eq:sep2} resembles the structure of the propagator in Eq.~\eqref{eq:propagator} and, in contrast to $\GCart$, prevents vanishing eigenvalues, since $|e^{\nu}|=e^{\operatorname{Re}\nu}>0$ ($\nu$ being the eigenvalues of the ``generator'' $\ln\Gop$). 

Nothing, however, keeps the eigenvalues \emph{on} the unit circle (Fig.~\ref{fig:genexp}a):
each $\Gop_{\theta_i}$ is derived from few noisy samples of a single projection, and taking the logarithm in Eq.~\eqref{eq:logsystem} amplifies this noise most severely along eigenvectors with small eigenvalues \cite{Higham2008}.

Finally, Eq.~\eqref{eq:Gtheta} assumes a constant step $\mathbf e_\theta$ along each
readout, which \emph{per se} restricts SC-GROG to ``radial-like'' trajectories.

\subsection{A Lie algebra approach for calibrating GRAPPA operators}\label{sec:ours}
In this work, an alternative way of estimating GRAPPA operators from non-Cartesian scans themselves is proposed. It exploits the linear regime of small k-space steps, which allows a single joint fit of the generators of $\Gop$.
\paragraph{The difference form.}
For consecutive readout samples with $\dk_n=\kv_{n+1}-\kv_n$, Eq.~\eqref{eq:gop} reads:
\begin{equation}
\begin{aligned}
  \sv_{n+1}&=\Gop(\dk_n)\,\sv_n
\end{aligned}
  \label{eq:perstep}
\end{equation}
Subtracting $\sv_n$ from both sides results in the discrete derivative of the coil signal along the readout:
\begin{equation}
\begin{aligned}
  \sv_{n+1}-\sv_n&=\big(\Gop(\dk_n)-I\big)\,\sv_n.
\end{aligned}
  \label{eq:diff0}
\end{equation}
In analogy to the time-evolution operator in Eq.~\eqref{eq:propagator}, an ideal GRAPPA operator would formally be driven by its generators $\Lop_a$
\begin{equation}
\begin{aligned}
  \Gop(\dk_n)=\exp\!\big(\sum_{a=x,y,z} \Delta k_{n,a}\Lop_a\big),
\end{aligned}
  \label{eq:badreg}
\end{equation}
one per axis, which are elements of the \emph{Lie algebra} of the group formed by $\Gop$ \cite{Hall}.
A Taylor expansion of the bracket on the right-hand side of Eq.~\eqref{eq:diff0} results in
\begin{equation}
\begin{aligned}
  &\quad \Gop(\dk_n)-I =\\
  & \qquad   \sum_{a=x,y,z} \Delta k_{n,a}\,\Lop_a +\tfrac12\Big(\sum_{a=x,y,z} \Delta k_{n,a}\,\Lop_a\Big)^2+\dots\\
  &   \; \; \approx\;\sum_{a=x,y,z} \Delta k_{n,a}\,\Lop_a,
\end{aligned}
  \label{eq:taylor}
\end{equation}
while the truncation performed in the last step delivers the first order approximation and corresponds to the tangent to the k-space orbit of an imaged object (Fig.~\ref{fig:genexp}b).
\paragraph{The joint fit.}
This assumption means leaving the curved manifold of $\Gop$ and ``traveling'' simply by the generators $\Lop_a$ themselves. In a regime of sufficiently small $\dk_n$, inserting the truncated version into Eq.~\eqref{eq:diff0} allows fitting $\Lop_a$ directly \emph{within} the Lie algebra, i.e. the tangent space at the identity:
\begin{equation}
  \;\sv_{n+1}-\sv_n\;\approx\;\Big(\sum_{a=x,y,z} \Delta k_{n,a}\,\Lop_a\Big)\sv_n\;
  \label{eq:diffform}
\end{equation}
In general, such pairs of a non-Cartesian trajectory have step components on every axis, thus a single pair models the combined effect of all three generators and not of a single isolated one.
As they enter Eq.~\eqref{eq:diffform} as a linear combination, however, pooling all pairs in one joint fit recovers all components of $\Lop_a$ from the many directions the trajectory holds \emph{at once}.

To this end, $M$ pairs of consecutive samples can be collected from the readouts of a specific non-Cartesian scan. In analogy to $S_0$ in Eq.~\eqref{eq:fitG}, their first samples $\sv_n$ are arranged as the columns of an $\NC \times M$ matrix $S$, and the differences $\sv_{n+1}-\sv_n$ of Eq.~\eqref{eq:diffform} as the columns of $\Delta S$, resulting in:
\begin{equation}
  \underbrace{\Delta S}_{\NC\times M}\;\approx\;\sum_{a=x,y,z}\underbrace{\Lop_a}_{\NC\times \NC}\,\underbrace{S}_{\NC\times M}\,\underbrace{D_a}_{M\times M},
  \label{eq:model}
\end{equation}
with $D_a=\operatorname{diag}(\Delta k_{1,a},\dots,\Delta k_{M,a})$ holding the step of every pair along axis $a$.
Substituting $\Phi=[\,S D_x;\,S D_y;\,S D_z\,]$ (dimension: $3\NC\times M$), Eq.~\eqref{eq:model} turns into
\begin{equation}
  \Delta S\;\approx\;[\Lop_x\,\Lop_y\,\Lop_z]\,\Phi,
  \label{eq:modelPhi}
\end{equation}
which is a linear system and can be solved for $[\Lop_x\,\Lop_y\,\Lop_z]$ in accordance with the Tikhonov-regularized least-squares inversion applied for calibrating $\GCart$:
\begin{equation}
\begin{aligned}
  &\underbrace{[\Lop_x\,\Lop_y\,\Lop_z]}_{\NC\times 3\NC}=\arg\min\;\big\|\Delta S-[\Lop_x\,\Lop_y\,\Lop_z]\,\Phi\big\|_F^2\\[2pt]
  &\qquad \qquad \qquad \qquad  +\;\beta\,\big\|[\Lop_x\,\Lop_y\,\Lop_z]\big\|_F^2 ,
\end{aligned}
  \label{eq:Lfit}
\end{equation}
where $\beta$ is a scale invariant regularization parameter $\beta = \beta_{\mathrm{rel}}\,\frac{\operatorname{tr}(\Phi\Phi^{H})}{3\NC}$ as in Eq.~\eqref{eq:fitGreg}, and $\beta_{\mathrm{rel}}$ was set to $10^{-1}$ for all experiments in this paper. 

Compared to $\GCart$ and SC-GROG, which directly fit (and regularize) the operators, this approach called LieGROG optimizes for the generators, and the penalty $\beta$ thus shrinks
the $\Lop_a$. An operator is moreover defined for a specific step only and combines multiplicatively, whereas the generators simply add (Eq.~\eqref{eq:diffform}).
This is why $\GCart$ requires the pure axis steps of a Cartesian reference and SC-GROG its two-step approach.

\paragraph{Unitary projection.}
As an optional extension, the operator as estimated by LieGROG can be additionally projected onto its closest unitary matrix:
\begin{equation}
  \Gop_a^{\,\mathrm{u}}\;=\;UV^{H},
  \qquad \Gop_a = U\Sigma V^{H},
  \label{eq:polar}
\end{equation}
by setting its singular values to one, and thereby also the absolute value of all eigenvalues (not their phases). This pragmatically reestablishes
the norm-preserving property of the ideal operator, which can be advantageous in terms of noise, but may also introduce a systematic error in $\Gop_a$,
depending on whether the deviation from the unit circle is caused by noise or by the encoding capacity of the array.

\paragraph{Applying a sub-cell shift.}
To ultimately grid the non-Cartesian data onto the regular grid, each of the operators $\Gop_a$ can be built using $\Gop_a=\exp(\Lop_a)$ and sub-cell shifts 
can be performed as in Eq.~\eqref{eq:frac}.
$S$ now denotes the $\NC\times M$ matrix of \emph{all} acquired samples, and $\delta_{a,j}$ the sub-cell shift of sample $j$ to its nearest
grid node along axis $a$. 
With $[\bm\Lambda_a^{\delta}]_{mj}=\lambda_{a,m}^{\delta_{a,j}}$ and $\Gop_a = \Em_a \bm\Lambda_a \Em_a^{-1}$, all samples can be shifted along axis $a$ \emph{at
once} by
\begin{equation}
  S \;\leftarrow\; \Em_a\,\big(\bm\Lambda_a^{\delta}\odot(\Em_a^{-1}S)\big),
  \label{eq:shiftall}
\end{equation}
where $\odot$ denotes the element-wise product, and the axes are processed one after the other. 
This requires one eigendecomposition per axis, two $\NC\times\NC$ products applied to the entire dataset, 
and a scalar exponent per sample and eigenvalue in between. No operator needs to be determined per sample.

In contrast to the calibration of $\GCart$ and SC-GROG, increased damping within Eq.~\eqref{eq:Lfit} now drives $\Gop_a$ to $I$ and its $\lambda_m$ towards unity.

All three GROG calibration methods share the same \emph{array floor} and differ only in the estimation error added to it. 
The pooled fit of LieGROG involves no logarithms of noisy data and regularizes towards $I$, which is expected to be favorable with respect to noise.
The price is a bias introduced by truncating higher orders in Eq.~\eqref{eq:taylor}, and this bias shrinks as consecutive readout samples move closer together.   
The closer the calibration pairs, however, the lower the SNR of their difference, since signals become more and more identical and noise stays constant when $\dk_n \rightarrow 0$. 
$\Phi$ holds the step of each pair in each column such that the close pairs also constrain Eq.~\eqref{eq:Lfit} the least and the solution features minimum variance.

The following experiments were performed to investigate which calibration approach proves favorable in different typical 2D and 3D scenarios.

\section{Methods}\label{sec:methods}

LieGROG was implemented for execution on a GPU using Python 3.10 and the packages NumPy 2.2.5 and PyTorch 2.7.0 (CUDA 12.6). Independent of the trajectory and throughout the experiments of this paper, only
pairs of consecutive readout samples with a step below $0.6$ cells along every axis were pooled to estimate $\Lop_a$, and the calibration was restricted to data from the high-SNR inner region of k-space (calibration radius $r<0.34\,N$ cells, $N$ being the first matrix dimension, i.e. $68\,\%$ of $k_{\max}$ on an isotropic grid).
The unitary projection was tested as an additional option.

Corresponding Python-versions of $\GCart$ and SC-GROG were implemented for comparison and share the same \emph{application} module with LieGROG, so that only the calibration of $\Gop$ differs.
The regularization parameter $\beta_{\mathrm{rel}}$ of $\GCart$ was set to $10^{-3}$ throughout, which parameter sweeps confirmed to be near optimal for our data. It is of comparable size to the default value of open implementations \cite{pygrappa}.
The matrix inversions in SC-GROG were not regularized. This is also in compliance with open implementations \cite{BART} and proved favorable in comparison to a $\beta_{\mathrm{rel}}>0$ on our data. 

TorchKbNufft \cite{Muckley2020} was used to provide efficient GPU-accelerated references for NUFFT-based reconstructions. Default parameters were used, in particular, a kernel size of $J=6$ and an oversampling factor of $\alpha = 2$.
The latter is not the decisive compute bottleneck and a reduction of it would weaken the quality of the benchmark reconstructions.

For undersampled non-Cartesian data, a CG-SENSE \cite{Pruessmann2001} algorithm \cite{Maier2021} was used. Applying LieGROG delivers a Cartesian k-space, i.e. Cartesian CG-SENSE can be used afterwards.
For comparison, Toeplitz-embedding \cite{Muckley2020} was used as normal operator in CG-SENSE, which represents a benchmark for efficient iterative parallel imaging. All CG-SENSE reconstructions used a regularization parameter of $\lambda = 5 \cdot 10^{-2}$, which proved a good compromise for all experiments. Since the normal operators of the different gridding routes feature different intrinsic scalings, the same $\lambda$ would still regularize corresponding CG-SENSE reconstructions differently. Therefore, before the actual reconstructions, the Cartesian normal operator (GROG based gridding) and the Toeplitz normal operator were applied to the same random test image, and $\lambda$ of the Toeplitz-based CG-SENSE iterations was multiplied by the ratio of the mean magnitudes of the results.
By this, all CG-SENSE reconstructions of this paper share a comparable regularization. Both routes were implemented with peak VRAM in mind: 
For LieGROG, the operator is applied to blocks of samples, while the NUFFT adjoint and both CG-SENSE variants, Toeplitz and
Cartesian, process the channels one after another in 3D and all at once in 2D.

Where necessary, coil sensitivities were estimated using ESPIRiT \cite{Uecker2014} and coil compression was performed as described in \cite{Huang2008}.
DCF were estimated by the TorchKbNufft-implementation of the algorithms proposed by Pipe and Menon \cite{PipeMenon1999}.

Software implementations were supported by a large language model based coding agent. Details are given in the Declaration on the Use of AI Technologies. Any generated source code was thoroughly reviewed.

All scripts were run on a PC with an Intel(R) Core(TM) i9-10850K CPU @ 3.60GHz with 128 GB memory equipped with an NVIDIA RTX A6000 ($48$\,GB) GPU.

\begin{figure}[!htb]\centering
  \includegraphics[width=\linewidth]{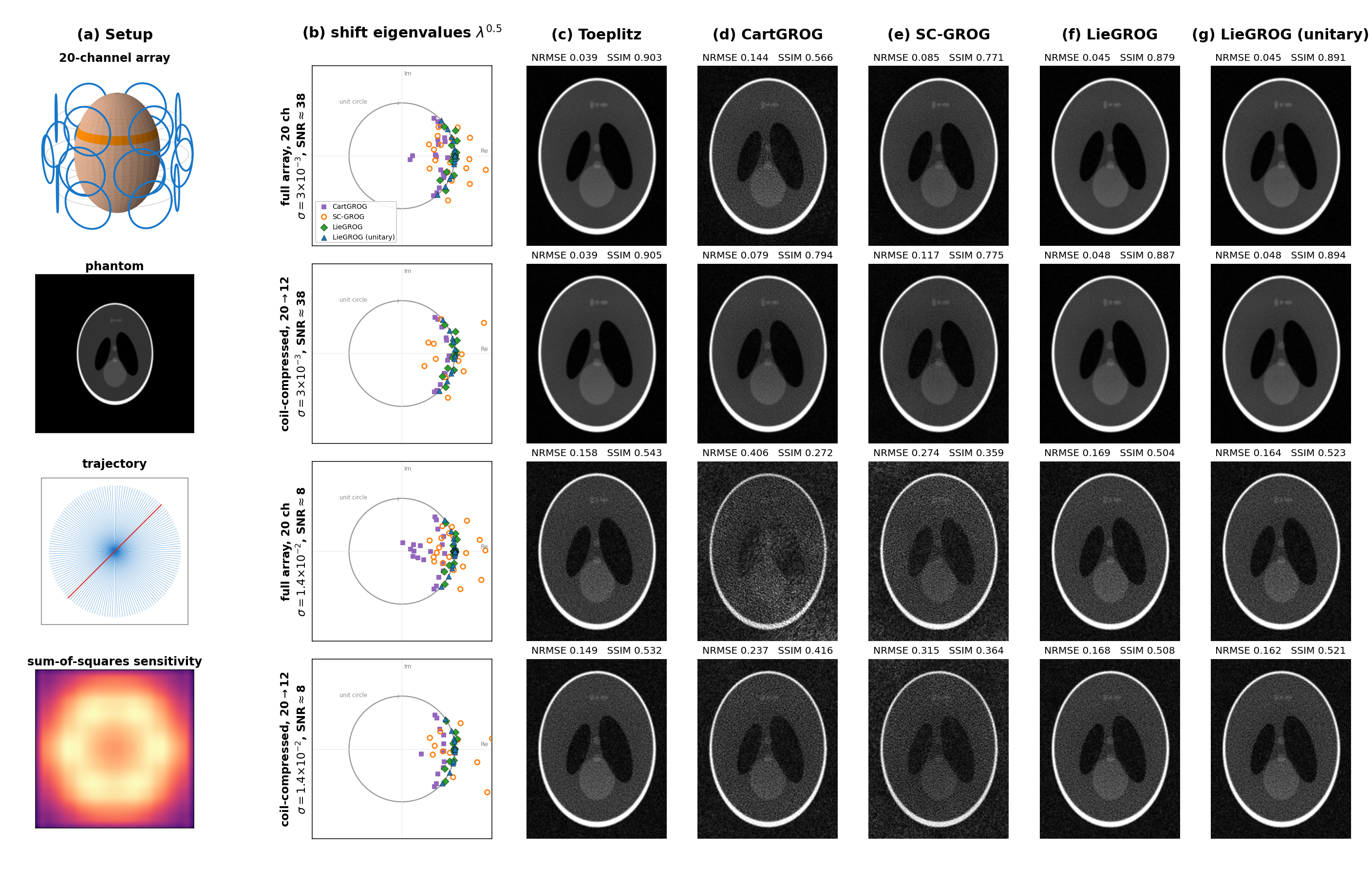}
  \caption{\textbf{Simulations in analytical phantom and exact k-space: Radial Trajectory.}
  The column in \textbf{(a)} shows a schematic view of the simulated fixed $20$-channel head array (three rings, $6{+}8{+}6$ surface loops)
  and the ``imaged'' slice depicting a Shepp--Logan phantom in orange. A radial sampling pattern was used ($160$ spokes, $256$ matrix). 
  In \textbf{(b)}, the shift eigenvalues to the power of $0.5$ are plotted for the different GRAPPA operators; $0.5$ being the largest shift distance.
  The following columns \textbf{(c-g)} show the results when reconstructing the data with Toeplitz-based CG-SENSE, as well as with Cartesian CG-SENSE 
  after transferring the radial data to a Cartesian grid by $\GCart$, SC-GROG, LieGROG and LieGROG with an additional unitary projection. Two different noise levels 
  were investigated, and the reconstructions were both performed for the full ($20$) and a compressed array with 12 virtual coils.}
  \label{fig:modes}
\end{figure}

\subsection{Analytic phantom and exact k-space}\label{sec:phantom}
To first evaluate LieGROG in a regime free from inverse crimes and ``hidden'' noise \cite{Wang2024}, \textit{in-silico} simulations based on a genuine ``ground truth'' were performed.
Analytical multi-coil phantom data were simulated for this purpose as proposed in \cite{GuerquinKern2012}. 
A two-dimensional Shepp--Logan phantom with closed-form Fourier transform \cite{GuerquinKern2012,SheppLogan1974} was combined with the band-limited sensitivities of a modelled physical $20$-channel head array.
With this setup, the signal at \emph{any} $\kv$ can be calculated exactly, without gridding and free of noise. 

Data on both an undersampled radial ($160$ spokes, full angular coverage) and an undersampled spiral 2D trajectory were simulated ($52$ equidistant interleaves, each a full spiral-out readout with gradient
rewind \cite{Schad2025}), together with the ground truth on the Cartesian grid points. 
Complex Gaussian noise was added to the simulated radial and spiral ``rawdata''. Noise standard deviations were chosen as $\sigma$ times the peak k-space magnitude:
$\sigma=3\times10^{-3}$ and $1.4\times10^{-2}$, resulting in an image SNR of ${\approx}38$ and ${\approx}8$ for radial, as well as ${\approx}44$ and ${\approx}10$ for spiral.  
Images were reconstructed on a $256 \times 256$ matrix using $15$ iterations of CG-SENSE. The exact coil sensitivities as simulated were used both for the full and a compressed array ($20\!\to\!12$).

The normalized root-mean-squared error (NRMSE) and the structural similarity index (SSIM) between the different reconstructions and the analytical phantom were calculated for the object support, after normalizing each image to its maximum. The exact reference was multiplied by the sum-of-squares profile of the coil sensitivities.

Furthermore, the deviations between the shifted data points in k-space with respect to the noise free analytical Cartesian equivalent were evaluated directly after the gridding procedures.

\begin{figure}[!htb]\centering
  \includegraphics[width=0.85\linewidth]{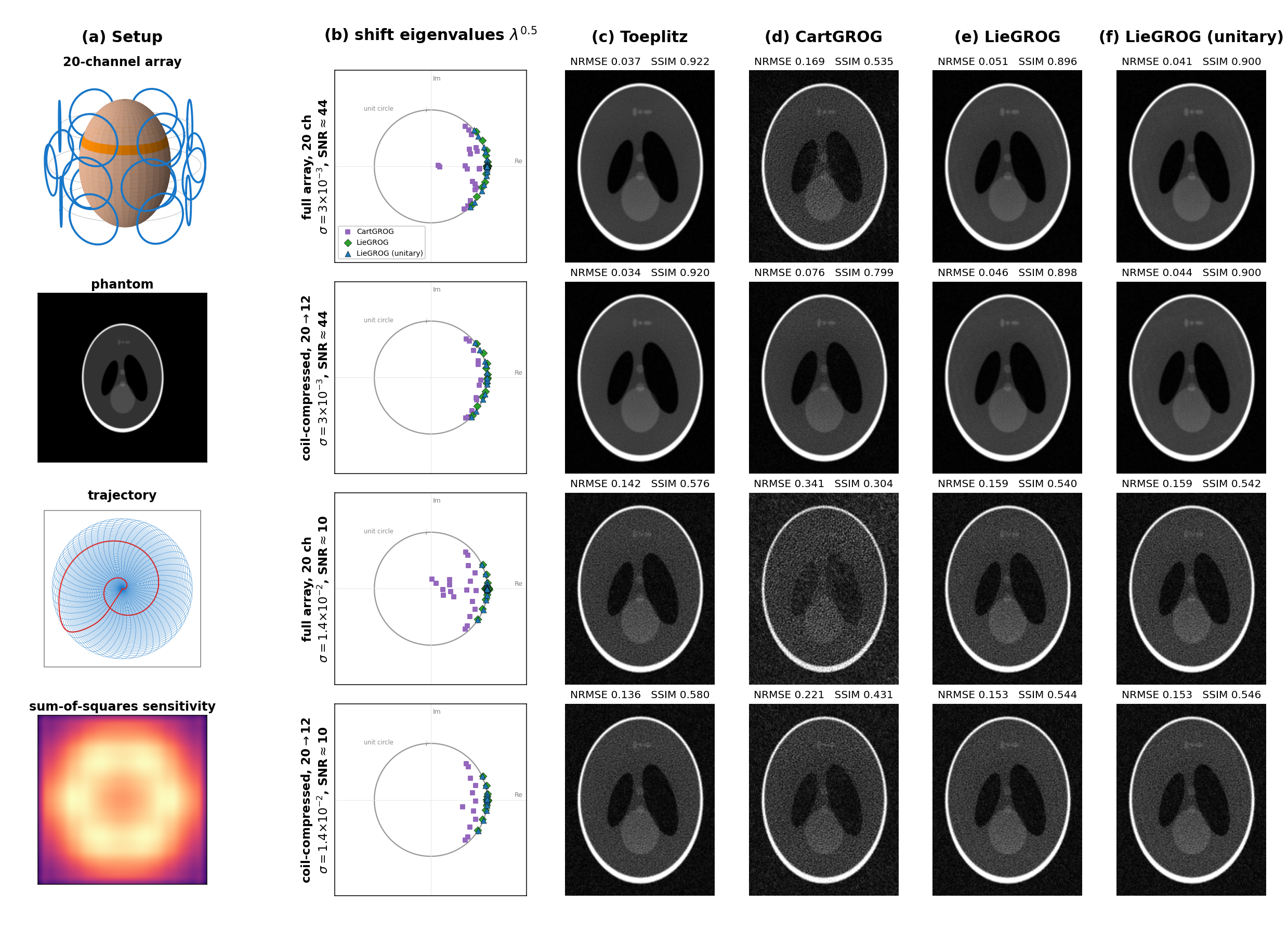}
  \caption{\textbf{Simulations in analytical phantom and exact k-space: Spiral Trajectory.}
  The column in \textbf{(a)} shows a schematic view of the simulated fixed $20$-channel head array and the ``imaged'' slice in orange. A $52$-interleave spiral was reconstructed on a $256 \times 256$ grid. 
  SC-GROG is omitted as its per-ray step needs constant-angular-velocity resampling and is undefined for a general spiral. In \textbf{(b)}, the shift eigenvalues to the power of $0.5$ are plotted for the different GRAPPA operators.
  The following columns \textbf{(c-f)} show the results when reconstructing the data with Toeplitz-based CG-SENSE, as well as with Cartesian CG-SENSE 
  after transferring the spiral data to a Cartesian grid by $\GCart$, LieGROG and LieGROG with an additional unitary projection. Two different noise levels 
  were investigated, and the reconstructions were both performed for the full ($20$) and a compressed array with 12 virtual coils.}
  \label{fig:modesspiral}
\end{figure}

\subsection{MR Experiments}\label{sec:mrexp}

LieGROG was then tested and evaluated in different non-Cartesian 2D and 3D MR-acquisitions.
In addition to a qualitative visual inspection, NUFFT/Toeplitz-based reconstructions were compared to the LieGROG approach by calculating the NRMSE between the two, 
thereby regarding the NUFFT/Toeplitz reconstruction as a pseudo-reference. For each dataset, coil sensitivities were estimated by applying ESPIRiT to a central part of a Cartesian
k-space as estimated by LieGROG. These sensitivities were shared between the different routes (NUFFT/Toeplitz \& LieGROG).

The overall and the step-wise run-times were measured together with the peak VRAM consumptions of each approach. Before any timing, a warm-up ``dummy''-call was performed
to not disadvantage the approach which was called first on the GPU.  

\paragraph{3D FLORET in a structure phantom}

An ultra-short echo time (UTE) pulse sequence based on a 3D Fermat looped, orthogonally encoded trajectory (FLORET, \cite{Pipe2011}) was implemented in pulseq \cite{Layton2017}
and applied on a 3T system (Magnetom Prisma, Siemens Healthineers, Forchheim, Germany) to scan a structure phantom (flip angle =  3°, repetition
time TR = 5\,ms, echo time TE = 40\,$\mu$s, $50{,}895$ arms in total, $980$ readout samples per arm, $1$\,mm isotropic resolution).
$26$ receive channels were compressed to $8$ virtual coils \cite{Huang2008} and data were gridded onto a $256^3$ matrix by both LieGROG and the NUFFT (with DCF). 
Coils were combined by coil sensitivity maps estimated by ESPIRiT \cite{Uecker2014}. In addition, the data were retrospectively
undersampled to $15\%$ of the arms and reconstructed by LieGROG followed by Cartesian CG-SENSE and Toeplitz-based CG-SENSE.

\begin{figure}[!htb]\centering
  \includegraphics[width=\linewidth]{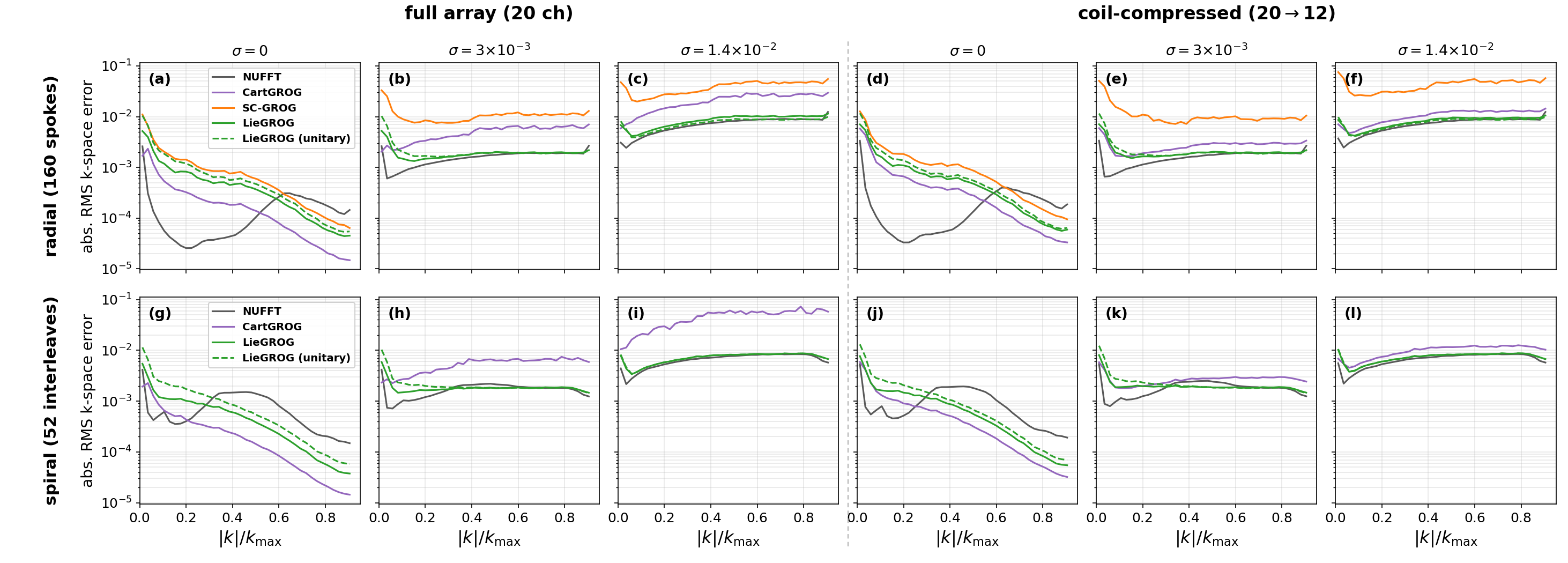}
  \caption{\textbf{k-space error.} The plots show the absolute RMSE of the shifted values --- directly after gridding --- against the exact analytic ground truth on the occupied cells of the Cartesian grid, for both radial and spiral.
  The errors were averaged within 40 k-space rings of equal width between the origin and $k_{\max}$. To compensate different scalings --- mainly due to the DCF --- the multi-coil data of each route were divided by a single complex gain,
  chosen such that the scaled ground truth best matches the gridded data in the least-squares sense. The NUFFT is represented by its density-compensated adjoint. Different noise levels (including $\sigma=0$) and both the full array and a compressed array with 12 virtual coils are shown. Dashed: LieGROG with unitary projection.}
  \label{fig:kspace}
\end{figure}

\paragraph{3D VASP in the thorax.}

Second, the thorax of a healthy volunteer was scanned \textit{in-vivo} using 3D radial imaging with spiral phyllotaxis (VASP \cite{Krishnamoorthy2021}) implemented in pulseq and on the same scanner. 
The study was approved by the local ethics committee (ID 204/23-sc) and written informed consent was obtained. $25{,}814$ radial projections with $330$ readout points each and $20$ receive channels coil-compressed to
$8$ were used (TR = 1.27 ms, TE = 40 $\mu$s, flip angle = 3°, $366\times274\times366$ grid). As this acquisition was already slightly undersampled, the data were reconstructed by CG-SENSE, as in the last part of the previous phantom experiment.
Data were then retrospectively undersampled to 20\% and the reconstructions were repeated.

\paragraph{3D IR-UTE of the knee.}

Next, a center-out radial UTE sequence (a.k.a. Koosh-Ball trajectory) was additionally equipped with an inversion pulse as proposed in \cite{Ma2020,Nunn2026} to selectively depict bony tissues of the knee with hyperintense contrast (IR-UTE).
The pulseq-based sequence was applied in a healthy volunteer (ethics approval ID 122/23-sc), again at the same 3T scanner. Successive inversions were spaced by 150 ms, each followed by a block of seven radial projections (TR = 3.8 ms, TE = 30 $\mu$s), 
centered at a TI of 64 ms, to null contributions from tissues with longer $T_2$. $84{,}000$ spokes were acquired with $320$ readout samples each (data were acquired for the study reported in \cite{Nunn2026}, see reference for more details). $15$ receive channels were compressed to $12$. To suppress the low-SNR end of the readouts, a Fermi roll-off (50\,\% at $86\,\%$ of the readout, width of $10$ samples) was applied to the raw data of both reconstruction routes. Images were reconstructed on a $384^3$ grid using $15$ CG-SENSE iterations.
Again, LieGROG followed by Cartesian CG-SENSE and Toeplitz-based CG-SENSE were used to reconstruct data.

\paragraph{2D spiral real-time cardiac cine.}
Finally, the spiral waveform as used in Sec.~\ref{sec:phantom} was used to perform real-time cardiac imaging as reported in \cite{Schad2025,Schad2026,Wech2025} (ethics approval ID 173/22\_skpm).
Data from a healthy volunteer, scanned on a 1.5T system (Magnetom Avanto, Siemens Healthineers, Forchheim, Germany) were used. Each frame was sampled by $13$ equidistant spiral
interleaves \cite{Schad2025}, with $30$ receive channels coil-compressed to $12$. After a single temporal-mean calibration, the frames were reconstructed one after another, both by LieGROG followed by Cartesian CG-SENSE and by Toeplitz-based CG-SENSE.

\begin{figure}[!htb]\centering
  \includegraphics[width=\linewidth]{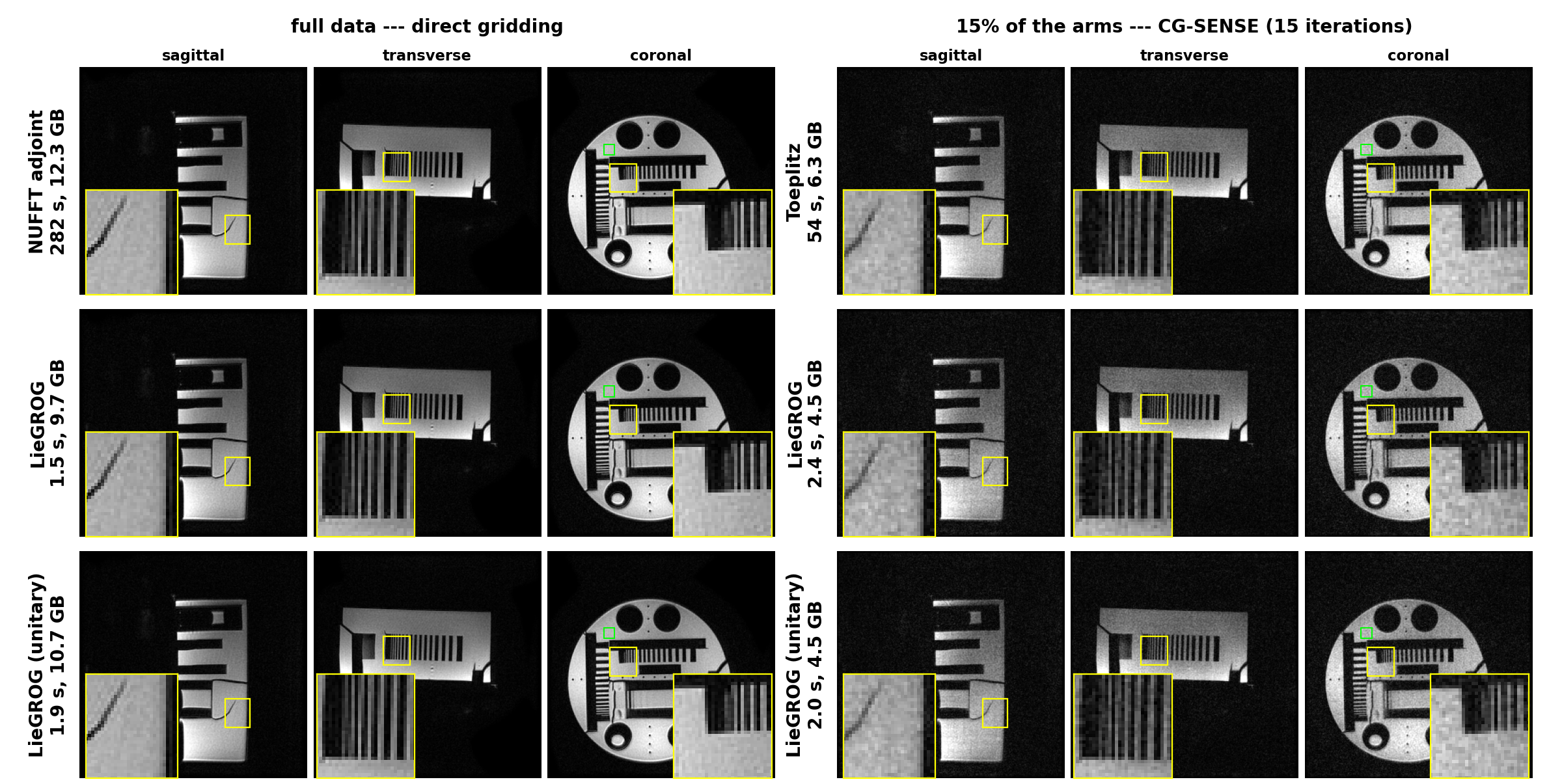}
  \caption{\textbf{3D FLORET in a structure phantom.}
  \emph{Left block, fully sampled:} density-compensated NUFFT adjoint vs. LieGROG vs. LieGROG + unitary projection (rows; columns show sagittal, transverse and coronal views). Coils combined using shared sensitivity maps,
  as determined by ESPIRiT. \emph{Right block, 15\% of the arms:} 15-iteration CG-SENSE with shared ESPIRiT maps, the top row using Toeplitz embedding.
  In each block, the ESPIRiT maps are estimated from the $48^3$ center of the corresponding LieGROG-gridded data and shared across the methods.
  Yellow insets zoom the crack (sagittal) and the fine resolution comb (transverse/coronal). SNR was estimated within the homogeneous region marked by the green box (five coronal slices). 
  For the full data, SNR was 23.4 for NUFFT and 21.9 for LieGROG. With an additional unitary projection, the same SNR (23.5) as in NUFFT was obtained.
  When using only 15\% of the data, SNR was 12.7 (Toeplitz) vs. 11.6 (LieGROG) vs. 12.2 (LieGROG unitary). LieGROG was clearly faster and required less peak VRAM.}
  \label{fig:floret}
\end{figure}

\section{Results}\label{sec:results}

\paragraph{Analytic phantom}\label{sec:results-insilico}
Fig.~\ref{fig:modes} shows the results of the experiments using analytical radial data. In addition to the setup in (a) and the reconstructions with the different approaches in (c-g), the $0.5$th powers of the eigenvalues of the respective GRAPPA operators are plotted in (b) (see Eq.~\eqref{eq:frac}; $0.5$ being the largest shift applied).
For the full array with 20 channels and high SNR (first row), $\GCart$ features eigenvalues with an absolute value $|\lambda_m^{0.5}|$ clearly smaller than $1$; two of them are close to zero. Those of SC-GROG scatter around $1$, while LieGROG delivers $|\lambda_m^{0.5}|$ close to $1$. The additional unitary projection draws the eigenvalues onto the unit circle.
The reconstruction of $\GCart$ is appearing as the most noisy one, followed by SC-GROG with moderate SNR. Toeplitz-based CG-SENSE and the LieGROG approaches appear similar with NRMSE and SSIM close together, slightly in favor of Toeplitz. When compressing coils to $12$ virtual ones, the $|\lambda_m^{0.5}|$ of $\GCart$ move closer to the unit circle
and the quality of the corresponding images improves, whereas the eigenvalues of SC-GROG scatter even more and its NRMSE increases. However, measures and also the visual appearance are still inferior to LieGROG and Toeplitz. The latter provides the lowest NRMSE and the highest SSIM, closely followed by LieGROG.
For the lower SNR, results are similar, while the noise amplification for the full array becomes intolerable for $\GCart$ and borderline for SC-GROG. Coil compression improves the results for $\GCart$, but, again, LieGROG (with and without additional unitary projection) clearly outperforms the other calibration techniques and closely follows Toeplitz.

The corresponding simulations with a spiral readout (see Fig.~\ref{fig:modesspiral}) did not match the requirements of SC-GROG, so this calibration method was not tested here. Apart from that, results are mostly similar with the radial trajectory: The uncompressed array results in very small eigenvalues for $\GCart$, which amplify the noise strongly for the backward shifts.
LieGROG is again robust and shows comparable performance to the Toeplitz benchmark.

The unitary projection leads to only minor differences in the resulting images, leaving NRMSE and SSIM equal or slightly improved in all cases. The largest difference was observed for the spiral trajectory with the full array at $\sigma=3\times10^{-3}$ (NRMSE $0.051$ vs. $0.042$).

Comparing the gridded values with the analytical ground truth on the Cartesian grid, as performed in Fig.~\ref{fig:kspace}, provides further insights. The discrepancy between the accuracy of the three GRAPPA calibration techniques can be clearly observed in this picture, too.
For $\sigma = 0$, no noise can be amplified and $\GCart$ clearly outperforms SC-GROG and LieGROG for both trajectories. Already for low noise, however, LieGROG has a clearly smaller RMS compared to $\GCart$ and SC-GROG. For high noise, a similar ranking results. $\GCart$ still has a lower error in k-space
than SC-GROG, even though the error of $\GCart$ seems to be amplified more strongly by CG-SENSE as can be seen in Figs.~\ref{fig:modes} and \ref{fig:modesspiral}. Coil compression mainly helps $\GCart$.
In the center of k-space, most pronounced in the noise-free run, the NUFFT estimates outperform the GRAPPA operator based approaches, of which $\GCart$ comes closest without noise and LieGROG as soon as noise is present. For higher k-values, the GRAPPA operator based approaches partly outperform the NUFFT in the noise-free case, and LieGROG approaches the NUFFT for growing noise. 

\begin{figure}[!htb]\centering
  \includegraphics[width=\linewidth]{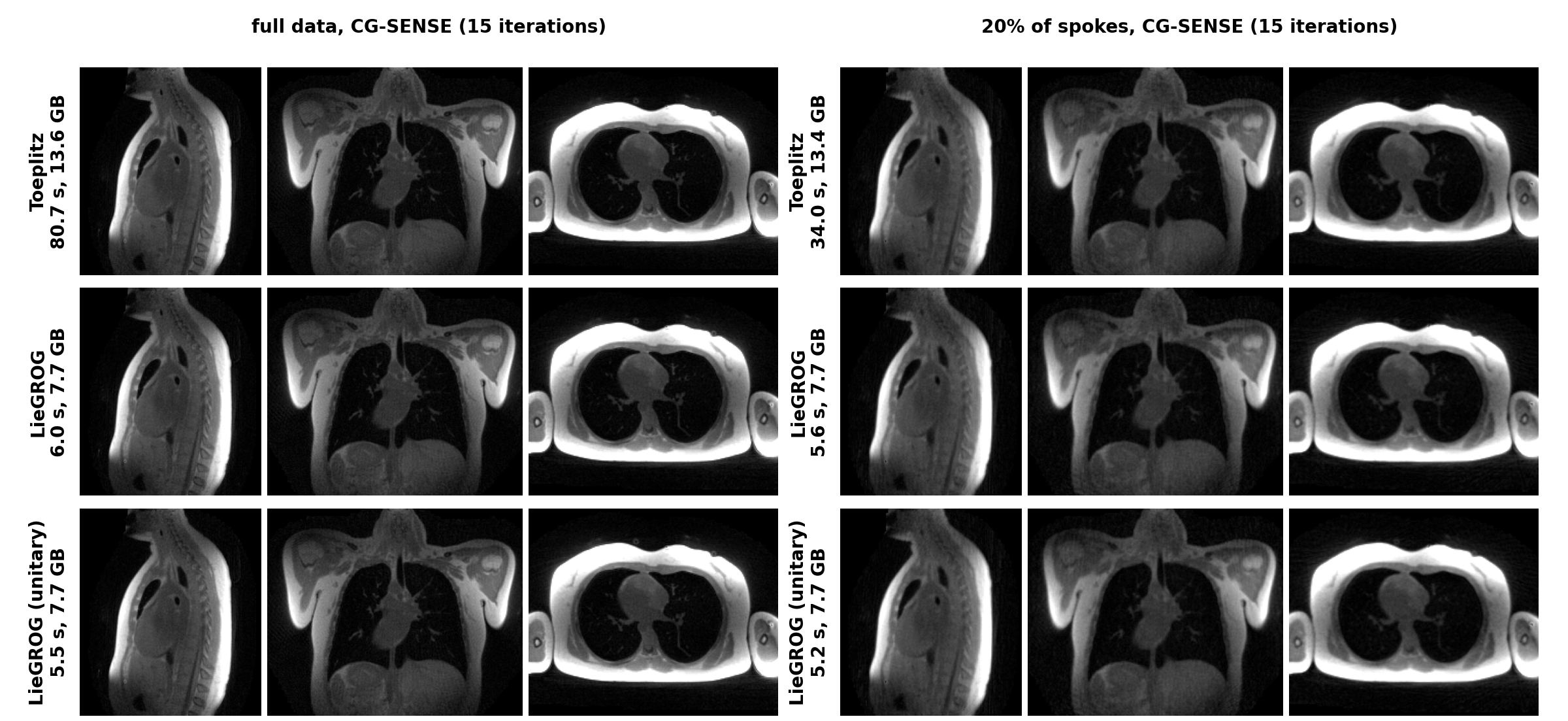}
  \caption{\textbf{3D anisotropic radial center-out (VASP) in the thorax of a healthy volunteer.} Both
  the full acquisition (left three columns) and $20\%$ of its projections (right three columns) are
  reconstructed with $15$ CG-SENSE iterations and per-level 3D-ESPIRiT maps. A bias field correction as in \cite{Axel1987} was applied. Rows
  are Toeplitz, LieGROG and LieGROG with unitary projection. LieGROG reconstructs $13\times$
  (full) and $6\times$ ($20\%$) faster and at roughly half the memory.}
  \label{fig:lung}
\end{figure}

\paragraph{3D FLORET in a structure phantom.}
Fig.~\ref{fig:floret} compares the results obtained by applying the adjoint NUFFT and LieGROG plus an inverse FFT to the fully sampled data (left block). On the right, Toeplitz-based CG-SENSE and LieGROG-gridding followed by Cartesian CG-SENSE
were used to reconstruct undersampled data. For both the full and the artificially accelerated data, reconstructions can barely be distinguished by the eye. 

The sharpness, as can be assessed by the insets providing a zoom into the comb structure and a crack within the material, appears
equivalent. The SNR of the NUFFT/Toeplitz reconstructions, as calculated for the green ROI, is slightly superior with respect to standard LieGROG and close (undersampled) to equivalent (full), 
when applying the additional unitary projection. (full data 23.4 (NUFFT) vs. 21.9 (LieGROG) vs. 23.5 (LieGROG + unitary); 15\% of data: 12.7 vs. 11.6 vs. 12.2).
The NRMSE with respect to NUFFT/Toeplitz was 0.049 (LieGROG) and 0.058 (LieGROG + unitary) for the full data, and 0.141 \& 0.132 for the undersampled set.

\begin{figure}[!htb]\centering
  \includegraphics[width=0.7\linewidth]{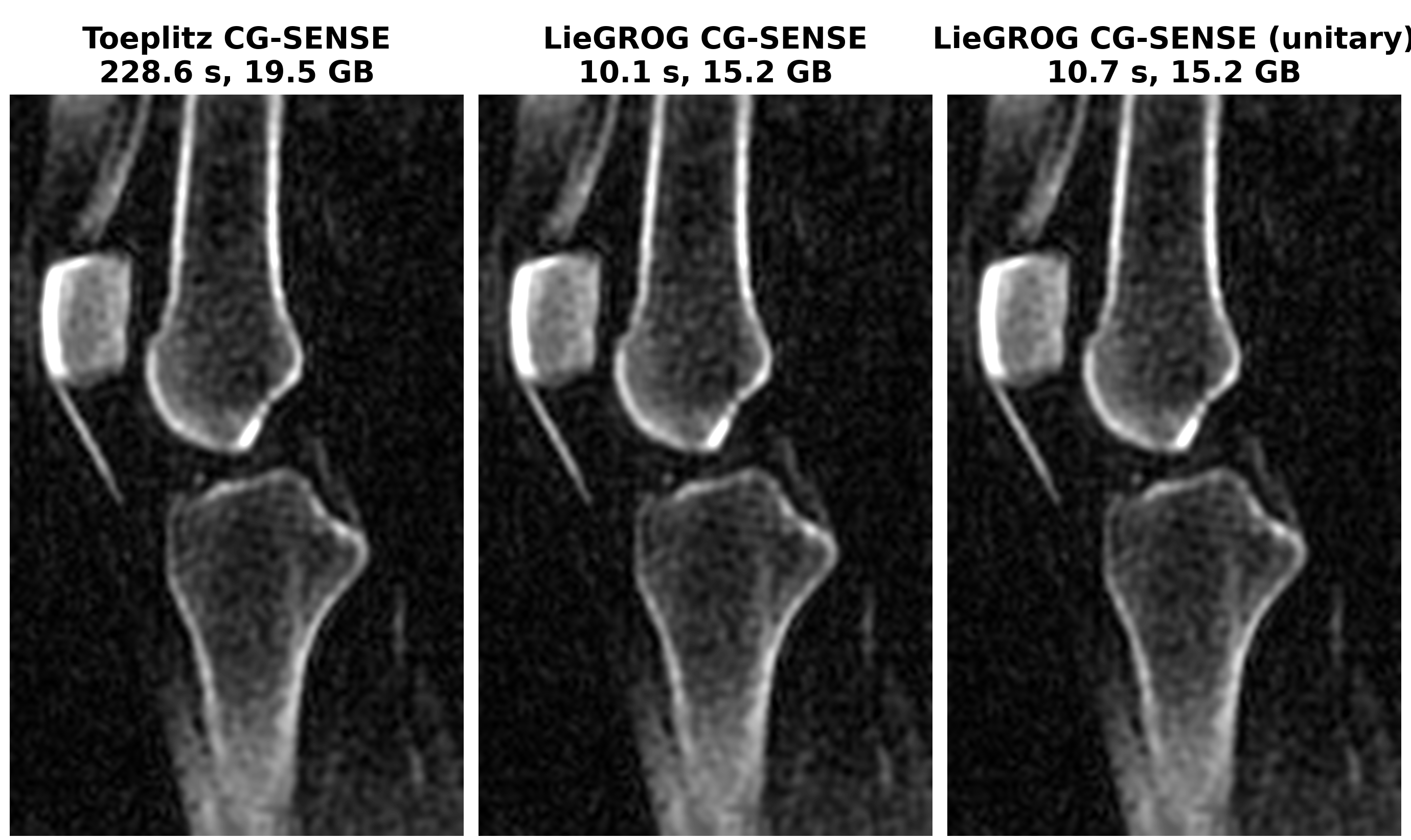}
  \caption{\textbf{3D IR-UTE of the knee.} All three reconstructions used $15$ CG-SENSE iterations.
  On the left, Toeplitz-based CG-SENSE is shown; the center image shows the result when applying LieGROG and subsequently CG-SENSE on the resulting Cartesian grid. On the right, 
  an additional unitary projection of the calibrated GRAPPA operator was used. Image quality is closely matched, while LieGROG is faster and needs less VRAM. The unitary projection shows no effect here.}
  \label{fig:knee}
\end{figure}

Determination of the shared ESPIRiT maps took $3.4$\,s for the full data and $4.9$\,s for the retrospectively downsampled version using $15\%$ of the data.
These times are excluded from the total times given in Fig.~\ref{fig:floret}. The adjoint NUFFT on full data took
$282.0$\,s, which is the sum of calculating the density compensation ($192.2$\,s), the adjoint ($89.7$\,s) and the coil combination
($<\!0.1$\,s). $12.3$\,GB peak GPU memory were measured. LieGROG took $1.5$\,s $=$
calibration $0.6$\,s $+$ gridding $0.8$\,s $+$ combination $0.1$\,s at $9.7$\,GB peak VRAM, and with the
unitary projection $1.9$\,s $=$ $1.0\,s+0.8\,s+0.1$\,s at $10.7$\,GB, i.e. $189\times$ and
$151\times$ faster at about $80\,\%$ of the memory. At $15\%$ of the arms (both $15$-iteration
CG-SENSE), the Toeplitz NUFFT route took $53.9$\,s ($=$ calculating the Toeplitz-kernel and the density compensation function
$36.0$\,s $+$ adjoint right-hand side $13.6$\,s $+$ solve $4.3$\,s) at $6.3$\,GB peak VRAM. LieGROG
required $2.4$\,s ($=$ $0.2\,s+0.1\,s+2.1$\,s) and $2.0$\,s with unitary projection ($=$ $0.1\,s+0.1\,s+1.7$\,s), both at
$4.5$\,GB, i.e. $22\times$ and $27\times$ faster. The trajectory pre-compute dominates the
NUFFT, although already the application of the adjoint NUFFT is clearly slower than the entire LieGROG chain. The same $15$ iterations run faster for LieGROG because
its normal operator is a plain Cartesian FFT rather than an oversampled Toeplitz convolution.

\paragraph{3D VASP in the thorax.}
LieGROG performed robustly also for the \textit{in-vivo} application in the thorax. In Fig.~\ref{fig:lung}, the three approaches are mostly equivalent in a visual inspection.
NRMSE with respect to NUFFT/Toeplitz were 0.041 \& 0.041 for LieGROG with and without unitary projection on the full data, and 0.043 \& 0.043 for 20\% of the MR data.

The estimation of ESPIRiT maps took $3.6$\,s and $3.7$\,s, for the full and the retrospectively undersampled data, respectively (excluded from the totals).
For the full acquisition, the Toeplitz NUFFT took $80.7$\,s ($=$ kernel compute including the
density compensation $43.6$\,s $+$ adjoint right-hand side $16.9$\,s $+$ solve $20.2$\,s) at
$13.6$\,GB. LieGROG lasted $6.0$\,s ($=$ calibration $0.4$\,s $+$ gridding $0.2$\,s $+$ solve
$5.4$\,s) and $5.5$\,s with additional unitary projection ($=$ $0.3\,s+0.2\,s+5.1$\,s). Peak VRAM consumption was $7.7$\,GB.
At $20\%$ of the projections, Toeplitz took $34.0$\,s $=$ $9.7\,s+4.1\,s+20.2$\,s at
$13.4$\,GB and LieGROG $5.6$\,s $=$ $0.1\,s+0.1\,s+5.4$\,s and the unitary branch $5.2$\,s $=$ $0.1\,s+0.1\,s+5.0$\,s at
$7.7$\,GB. LieGROG thus reconstructed about 13 times faster on the full data and about 6 times at $20\%$. It required roughly half the memory.

\paragraph{3D IR-UTE of the knee.}\label{sec:knee}
Comparable findings were made for the low-SNR regime of selective bone imaging (see Fig.~\ref{fig:knee}). The reconstruction quality is mostly equivalent. NRMSE was 0.070 for both LieGROG methods.

Reconstructing with Toeplitz-based CG-SENSE took $228.6$\,s ($=$ kernel including the iterative density compensation $130.6$\,s $+$
adjoint right-hand side $73.8$\,s $+$ solve $24.2$\,s) at $19.5$\,GB peak GPU memory.
LieGROG and the subsequent Cartesian CG-SENSE solve took $10.1$\,s ($=$ calibration $1.0$\,s $+$ gridding $0.7$\,s $+$ solve
$8.4$\,s) at $15.2$\,GB peak VRAM. LieGROG with the unitary projection showed again almost the same measures ($10.7$\,s $=$
$1.2+0.7+8.8$\,s, $15.2$\,GB peak VRAM). The shared ESPIRiT maps ($10.3$\,s) are excluded
from all three. The LieGROG route is about $23\times$ faster.

\begin{figure}[!htb]\centering
  \includegraphics[width=0.5\linewidth]{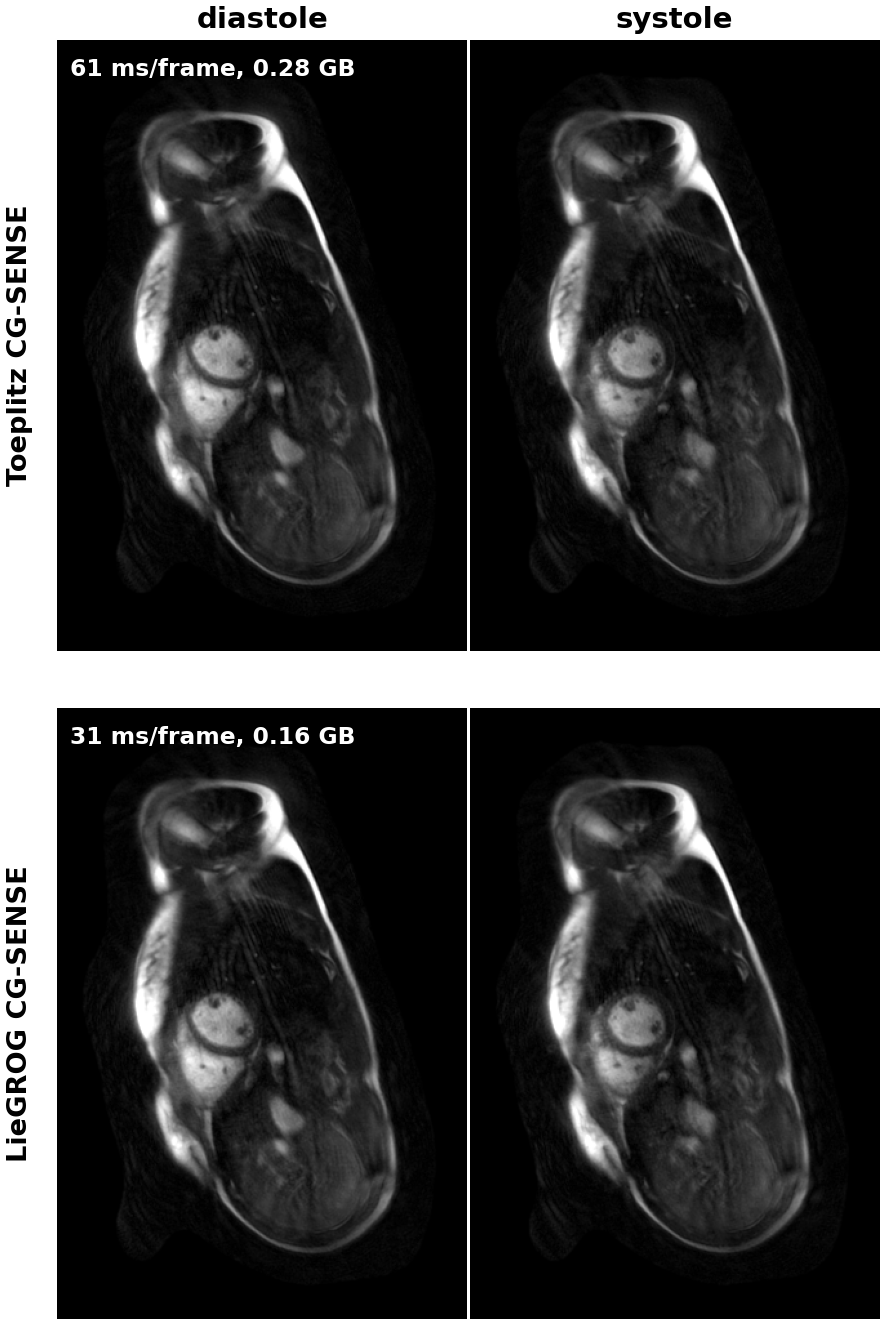}
  \caption{\textbf{2D Real-time cardiac cine.} CG-SENSE reconstructions (15 iterations) of real-time frames from a 2D spiral real-time sequence with a temporal footprint of $48$\,ms.
  End-diastole and end-systole, for Toeplitz and LieGROG. The given times exclude any pre-computes like estimation of coil sensitivities, Toeplitz-kernel, DCF and the GRAPPA operator. Dynamic version can be found in Supporting Information Video~S1.}
  \label{fig:rtheart}
\end{figure}

\paragraph{2D spiral real-time cardiac cine.}
Fig.~\ref{fig:rtheart} shows an end-diastolic and end-systolic frame of the real-time series, both for the Toeplitz-based and the LieGROG prepared CG-SENSE reconstruction.
The reconstructions appear identical and both show residual undersampling which parallel imaging cannot remove without further model- or data-driven regularization \cite{Schad2025,Schad2026,Wech2025}.
A complementary video of the cine can be found in the Supporting Information Video S1.

After a single calibration on the temporal mean ($0.03$\,s for the GRAPPA operators, $0.16$\,s for
the ESPIRiT maps, both needed once for the scanned slice), LieGROG plus CG-SENSE reconstructs at
$31$\,ms/frame, which corresponds to $\sim$$33$\,fps. The Toeplitz approach needs a trajectory-specific pre-compute for every
frame, which can only be processed in advance if the trajectory is known before the scan.
Including the density compensation, it took $226$\,ms (the per-frame run time in Fig.~\ref{fig:rtheart} excludes this).
The adjoint then took $17$\,ms and the CG-SENSE run $44$\,ms. Peak GPU memory was $0.16$\,GB per frame for LieGROG
and $0.28$\,GB for the NUFFT. The two agree to an NRMSE of $0.030$.

\section{Discussion}\label{sec:discussion}

\paragraph{Summary of findings}
LieGROG's approach of calibrating the GRAPPA operators by jointly estimating its generators in the Lie algebra delivered a
reliable and robust gridding tool for the ``real-world'' scenarios tested in this paper.
In particular, it clearly outperformed $\GCart$ and SC-GROG, as soon as the data were corrupted by a small amount of noise. Free of noise,
the regularization of the fit and the truncation error of LieGROG (see Eq.~\eqref{eq:taylor}) still rendered its accuracy inferior to the direct fit of $\GCart$, which in this regime only carries the \emph{array floor} (see Fig.~\ref{fig:kspace}).
Already for $\sigma=3\times10^{-3}$, however, the noise amplification of $\GCart$ is so strong that it lifts the k-space error clearly above LieGROG. SC-GROG amplifies noise in k-space even more, while the CG-SENSE routine applied subsequently propagates the errors
of the $\GCart$-gridded samples more strongly into the images (see Figs.~\ref{fig:modes} and \ref{fig:modesspiral}).

For the 2D radial and 2D spiral trajectory, the kernel-based interpolation of the densely sampled data in the center of k-space leads to superior gridding performance of NUFFT. 
Mainly the noise-free example in Fig.~\ref{fig:kspace} reveals the remaining model error of the GROG approaches (the \emph{array floor} and additional estimation errors).
For higher k-values, where the sampling is sparse, NUFFT does not profit from this any more and the GROG-methods outperform it in the noise-free case. 

For noisy (i.e. realistic) scenarios, LieGROG showed a largely comparable performance to NUFFT. Thus, its compromise between the different errors within the calibration (see Sec.~\ref{sec:ours}) seems to be clearly superior compared to $\GCart$ and SC-GROG here.

LieGROG furthermore outperformed NUFFT and Toeplitz-based reconstructions in terms of reconstruction speed across the diverse 2D and 3D examples investigated, and in terms of peak VRAM consumption in all 3D examples, while it was closely matching the image quality throughout.
In addition, LieGROG (like the other GROG approaches) intrinsically corrects for sampling density such that no DCF needs to be estimated.

Overall, LieGROG allowed transferring non-Cartesian data to a Cartesian grid in a simple, robust and fast manner. 
Any subsequent reconstruction steps or procedures could be performed in a strictly Cartesian manner; in particular based on straightforward forward and inverse FFTs.

\paragraph{Regularization parameters}
The way the different optimizations of this work were regularized had an impact on the findings reported. For reference methods, 
these were chosen as much as possible in line with reference implementations and to optimize benchmark performance.
In particular, Tikhonov regularization was not used for determining the inverse within the first step of SC-GROG. 
This would have worsened the noise amplification in our experiments. 

For $\GCart$, $\beta$ balances two aspects: it has to be large enough to keep the noise of the calibration data out of the singular directions of the inverse, and small enough to not bias the $\lambda_m$ in the range, where it inflates noise in backward shifts for gridding. 
For our experiments, the parameter sweeps mentioned in Sec.~\ref{sec:methods} resulted in an ``optimal'' $\beta_{\mathrm{rel}}=10^{-3}$.

Also, the regularization of the LieGROG-fit in Eq.~\eqref{eq:Lfit}, i.e. $\beta_{\mathrm{rel}} = 10^{-1}$, was chosen as a best compromise 
for the examples carrying any noise shown in this paper. A lower value would have been favorable for the performance on noise free data. When choosing $\beta_{\mathrm{rel}} = 10^{-4}$ the course for LieGROG is even slightly below $\GCart$ in Fig.~\ref{fig:kspace}(g) (data not shown).
The chosen $\beta_{\mathrm{rel}}$ ($10^{-1}$) proved however favorable and stable for the diverse (realistic) noise levels presented in this work.
The trajectories investigated here, i.e. 2D spirals, 3D FLORET, 3D Koosh-Ball, and similar approaches feature ramp sampling in the beginning of the readout and thus cover the range from small to large steps smoothly, 
which suits the weighting of the fit by $\Phi$.
In some scenarios the unitary projection can be favorable with respect to a lower noise amplification. As the latter is already low for plain LieGROG and the chosen regularization, it is regarded as a complementary option.
In this work, the additional projection proved to be beneficial for the SNR in the phantom scan using FLORET and for the NRMSE measured for the spiral simulation with the full array at moderate noise.

Finally, to allow a fair comparison of reconstruction performance and run-times between the LieGROG- and NUFFT/Toeplitz-driven pipelines, 15 iterations of CG-SENSE were used throughout and no additional termination criterion was set.

\paragraph{Execution Speed and VRAM Consumption}
Excluding the shared ESPIRiT calibration and including the determination of both the GRAPPA operator and the DCF and Toeplitz kernels, 
LieGROG-based CG-SENSE was faster than Toeplitz CG-SENSE by factors of 6 to 23 in the 3D experiments (e.g. 10.1 vs. 228.6\,s for the knee).
For the fully sampled FLORET data, LieGROG gridding was 189 times faster than the density-compensated NUFFT adjoint (1.5 vs. 282.0\,s, of which 192.2\,s are the density compensation and 89.7\,s the adjoint itself).
The adjoint was implemented to process the channels one at a time. A variant transforming all channels at once needed 38.5\,s instead of 89.7\,s, but 28.0 instead of 12.3\,GB.
In 3D, LieGROG required 1.3 to 1.8 times less peak GPU memory (e.g. 9.7 vs. 12.3\,GB for the full FLORET data and 7.7 vs. 13.6\,GB for the lung).
For 2D, peak VRAM was small for both approaches (0.16 vs. 0.28\,GB per cine frame).

The calibration of LieGROG represents a \emph{single} and fast least-squares fit (Eq.~\eqref{eq:Lfit}) across all selected signal pairs, 
which comprises two matrix products with $\Phi$ and one inversion of a $3\NC\times3\NC$ matrix.
It scales with the number of selected pairs and not with the matrix size.
For applying the shifts, no individual operators need to be calculated for each shift separately. Eq.~\eqref{eq:shiftall} requires one eigendecomposition per axis 
and two $\NC\times\NC$ products for the entire dataset. The individual shifts are taken into account as computationally cheap powers of the scalar eigenvalues, 
which can be efficiently broadcast on the GPU. The resulting matrix only needs to be multiplied element-wise with the data in the eigenbasis, such that no matrix power 
needs to be taken for any sample. Any subsequent reconstruction procedure (e.g. CG-SENSE) can be based on plain FFTs.

For real-time cardiac imaging, it was assumed that all calibrations (GRAPPA operator, Toeplitz kernel, DCF) can be determined beforehand. 
The reconstruction times per frame measured for LieGROG would in principle (not tested) allow ``real-time'' online depiction of the dynamic series.
With only $31$\,ms for shifting the samples and applying CG-SENSE, the total processing time is still below the acquisition footprint of $48$\,ms.
The Toeplitz approach needed $61$\,ms per frame and thus exceeded the footprint, even with its kernel determined beforehand. Including the kernel, it took $287$\,ms.
LieGROG can calibrate the GRAPPA operators prior to the acquisition if only some calibration data are available. The Toeplitz kernel can only be pre-estimated if the exact frame-wise trajectories are known in advance. 

An additional data-driven regularization as used in Variational Networks \cite{Hammernik2018} or similar approaches could remove the residual undersampling artifacts left by CG-SENSE without exceeding the tight time limits of real-time imaging.

\paragraph{Relation to other GROG methods}
In the experiments of this work, LieGROG proved to be clearly more robust to noise in comparison to the classical calibration techniques $\GCart$ \cite{Seiberlich2007} and SC-GROG \cite{Seiberlich2008},
where noise was either amplified by low eigenvalues or by taking the logarithm of angle-specific operators, which were fitted from a single noisy readout. 
This advantage of LieGROG clearly outweighed the systematic deviation introduced by neglecting higher orders of the Taylor expansion in Eq.~\eqref{eq:taylor}.

The logarithm in Eq.~\eqref{eq:logsystem} comes with two further issues for SC-GROG, which were not separately investigated in this work: 
the second step in Eq.~\eqref{eq:sep} fits data whose variances were made unequal by the logarithm, i.e. the unweighted solution is not of minimum variance, 
and eigenvalues with a phase close to $\pm\pi$ may be assigned to the wrong branch when data are noisy. LieGROG is free of both: it takes no logarithm of noisy data, and
the weighting by $\Phi$ in Eq.~\eqref{eq:Lfit} makes its fit one of minimum variance. 

More recently, a fundamentally different GROG procedure was presented, which estimates each Cartesian sample from several neighboring samples and models the corresponding 
GRAPPA kernels by an implicit neural network \cite{iGROG}. It is trained in a self-supervised fashion
and can also be favorable with respect to the classical methods in terms of noise amplification. 
It, however, requires a separate or synthesized calibration region and a complete network training.

\paragraph{Limitations}\label{sec:limitations}
As for $\GCart$ and SC-GROG, trajectories with a limited angle coverage for a specific direction only result in an ill-conditioned axis separation
and the NUFFT is clearly the better choice here. While an isolated scan of specific segments of k-space is rather uncommon, a varying contrast for the different readouts
that make up the k-space is typical for newer imaging approaches like magnetic resonance fingerprinting \cite{Ma2013}. Because only neighboring pairs of the same readout
enter the fit for LieGROG, its calibration should stay clearly less sensitive to contrast changes \emph{between} readouts than that of $\GCart$, which is calibrated on a Cartesian region synthesized from all readouts.

The GRAPPA operator is a characteristic of the individual investigation setup, i.e. the arrangement of the coils on the patient and its positioning.
$\Gop$ can thus not be pre-computed and needs to be determined for each exam. Its estimation is, however, cheap: Even for 3D, fitting the LieGROG operators 
took at most 1.2\,s. These can then be used as long as the arrangement of coils does not change.

Finally, LieGROG shares the same issues of gradient imperfections with any other approach for gridding non-Cartesian data. While the gradient system frequency response function or
gradient impulse response function represent generic tools to pre- or post-compensate for associated errors \cite{Stich2018, Scholten2023}, GROG itself was used to iteratively shift
readouts towards their actual position in k-space, as set by the potentially distorted gradient waveform \cite{Wech2015}.

\section{Conclusion}
In LieGROG, the group structure of the GRAPPA operators ($\Gop_x$, $\Gop_y$, $\Gop_z$) is exploited to calibrate them by a single joint
least-squares fit of their generators in the Lie algebra. This substantially reduces the noise amplification of the established calibration methods,
which may have prevented a broader clinical use of GROG so far.
LieGROG requires no additional reference and showed reliable performance across various noise levels and trajectories, including 2D radial, 2D variable density spiral, 3D radial (center-out) and 3D FLORET,
with no need for scan-specific adjustments. While providing closely matching image quality, LieGROG was clearly faster than NUFFT- or Toeplitz-based reference approaches for image reconstruction, and its
peak VRAM consumption was lower in all 3D experiments. By transferring non-Cartesian data to a Cartesian grid once, in a fast and straightforward manner, LieGROG makes the 
sampling advantages of non-Cartesian MRI (motion robustness, efficient scanning, etc.) accessible at the cost and simplicity of Cartesian reconstruction. 
Advanced (iterative) methods which were originally developed for Cartesian readouts can be easily extended to almost any non-Cartesian scan by simply adding
the LieGROG module.

\section*{Acknowledgement}
The author thanks Herbert Köstler for critical reading of the manuscript at multiple stages and for many valuable discussions that improved this work. The author further thanks Sebastian Scheidel, Philipp Nunn and Oliver Schad for the support with test data.

This work is dedicated to the memory of our dear friend Michael Kratochwil.

\section*{Conflict of Interest}
The author is named as inventor on a pending patent application related to the method described in this work, filed by the University of W\"urzburg.

\section*{Declaration on the Use of AI Technologies}
Anthropic Claude models (Claude Opus 4.8 and 5, Claude Fable 5 and 5.1), accessed through Claude Code, were used as a coding agent for implementing the methods, the experiment scripts and the figures. 
During the revision of the manuscript, the models were used for language editing and for proposing alternative phrasings of individual sentences, which were adopted in original or adapted form, 
as well as for validating numbers and equations against the code and its logs. 
All generated content was rigorously checked and the author takes full responsibility for the accuracy and integrity of the publication. 
All claims and statements expressed herein represent the author's own scientific positions.

\end{document}